# On the Use of Approximations in Statistical Physics

Conrado Hoffmann (1, 2)

((1) Laboratorio de Física del Sólido, Dto. de Física, Facultad de Cs. Exactas y Tecnología, Universidad Nacional de Tucumán. (2) Facultad Regional Tucumán, Universidad Tecnológica Nacional.)

**Abstract**

Two approximations are frequently used in statistical physics: the first one, which we shall name the *mean values approximation*, is generally (and improperly) named as "maximum term approximation". The second is the "Stirling approximation". In this paper we demonstrate that the error introduced by the first approximation is exactly compensated by the second approximation in the calculation of mean values of multinomial distributions.

**I-Introduction. Use of approximations.**

In the traditional formulation of statistical thermodynamics it is of common use the application of two approximations: the first one is the improperly named "maximum term approximation", and is customarily followed by the "Stirling approximation". We consider that the first approximation is improperly named bacause the procedure it prescribes is just to replace the number of occurrences of the events of interest, $n_i$, by its mean value, $\bar{n}_i$.

In this note we show that the mentioned approximations has the remarkable property of exactly compensate the errors introduced by each other, irrespectevely of the value of the numbers $n_i$, in the calculation of the mean value of random variables.

**II- Approximation using the mean values of $n_i$.**

Let us consider the random variable X wich can assume the value $x_1, x_2,...x_R$, with probabilities $p_1, p_2,...p_R$.. Its mean value is:

$$m_1 = p_1 x_1 + p_2 x_2 + \ldots + p_R x_R \qquad (1)$$

Let us consider now the random variable $Y_N$, sum of N independent random variables with the same probability distribution as X. Its first moment will be:

$$Y_N = N\, m_1. \qquad (2)$$

If we now calculate the first moment $Y_N$ using the multinomial probability of the occurrence of $n_1$ results $x_1$, $n_2$ results $x_2$, etc, we obtain the following expression:

$$\overline{Y_N} = \sum_{\{n_i\}} \frac{N!}{n_1! n_2! \ldots n_R!} p_1^{n_1} p_2^{n_2} \ldots p_R^{n_R} (n_1 x_1 + n_2 x_2 + \ldots n_R x_R) \qquad (3)$$

The "maximum term approximation" supposes that this sum may be approximated by one term in which the variables $n_i$ are replaced by their mean values, $\overline{n_i} = p_i\, N$. We obtain then the first approximation for $\overline{Y}_N$.

$$A_1 = \frac{N!}{p_1 N!\, p_2 N! \ldots p_R N!} p_1^{p_1 N} p_2^{p_2 N} \ldots p_R^{p_R N} (p_1 N x_1 + p_2 N x_2 + \ldots p_R N x_R) \qquad (4)$$

**III-The Stirling approximation**

If we now apply on expression (4) the approximation ln (s!) = s ln s – s we obtain for the second approximation to $\overline{Y}_N$, $A_2$, the following equation:

$$A_2 = \frac{N^N}{(p_1N)^{p_1N}(p_2N)^{p_2N}......(p_RN)^{p_RN}} p_1^{p_1N} p_2^{p_2N}......p_R^{p_RN} \times$$

$$\times N(p_1x_1 + p_2x_2 + ......p_RNx_R) \quad (5)$$

That is to say, we obtain again the exact value (note that the same result would be obtained if we use simply ln (s!) = s ln s).

We have demonstrated the following theorem: "The approximation ln (s!) = s ln s compensates exactly the errors introduced by the approximation that uses the mean values"